\documentclass[twocolumn,showpacs,amsmath,amssymb,prl]{revtex4}
\usepackage{graphicx}
\usepackage{dcolumn}
\usepackage{bm}
\begin{document}
\title{Frank's constant in the hexatic phase}
\author{P. Keim, G. Maret*, and H.H. von Gr\"unberg}
\affiliation{Universit\"at Graz, 8010 Graz, Austria\\
*Universit\"at Konstanz, 78457 Konstanz, Germany}
\date{\today}
\begin{abstract}
Using video-microscopy data of a two-dimensional colloidal system
the bond-order correlation function $G_6$ is calculated and used
to determine the temperature-dependence of both the orientational
correlation length $\xi_{6}$ in the isotropic liquid phase and the
Frank constant $F_A$ in the hexatic phase. $F_A$ takes the value
$72/\pi$ at the hexatic $\leftrightarrow$ isotropic liquid phase
transition and diverges at the hexatic $\leftrightarrow$ crystal
transition as predicted by the KTHNY-theory. This is a
quantitative test of the mechanism of breaking the orientational
symmetry by disclination unbinding.
\end{abstract}
 \maketitle

The theory of melting in two dimensions (2d) developed by
Kosterlitz, Thouless, Halperin, Nelson and Young (KTHNY-theory)
suggests a two-stage melting from the crystalline phase to the
isotropic liquid. The first transition at temperature $T_m$ is
driven by the dissociation of thermally activated dislocation
pairs into isolated dislocations breaking the translational
symmetry \cite{Kosterlitz73,Young79}. The fluid phase directly
above $T_m$ still exhibits orientational symmetry and is called
the hexatic phase. It may be viewed as an anisotropic fluid with a
six-fold director \cite{Halperin78,Nelson79} which is
characterized by a finite value of Frank's constant $F_A$, the
elastic modulus quantifying the orientational stiffness. At the
second transition at $T_i > T_m$, the dissociation of some of the
dislocations into free disclinations destroys the orientational
symmetry. Now, the fluid shows ordinary short-range rotational and
positional order as it is characteristic of an isotropic liquid.

Following an argument given in \cite{Kosterlitz73,Nelson79}, $T_m$
and $T_i$ can be estimated using the defect interaction
Hamiltonian $H_{d}$ between a pair of disclinations ($d=disc$) and
a pair of dislocations ($d=disl$) which for both defect pairs and
at large distances goes like $H_{d} \sim c_d \log r$ with the
dimensionless strength parameter $c_d$ depending on the defect
type. Defect dissociation is completed at a temperature where the
thermally averaged pair distance $\langle r^2_{d} \rangle$
diverges. Evaluating this expression for $H_d$ one generally finds
divergence if $c_d = 4$. The unbinding condition $c_d = 4$
translates into $\lim_{T\to T_m^-}\beta\mathcal{K}(T)a_0^2 =
16\pi$ for dislocation pairs ($\beta=1/k_BT$, $a_0$ is lattice
spacing) and into $\lim_{T\to T_i^-}\beta F_A(T) = 72/\pi$ for
disclination pairs, where $\mathcal{K}$ is the Young's modulus of
the crystal. Connecting thus the defect pair unbinding condition
to the two transition temperatures $T_i$ and $T_m$, two
expressions are obtained that summarize the microscopic
explanation of the KTHNY theory for two-stage melting.

In this Letter we study the temperature-dependence of Frank's
constant of a 2D system in the hexatic phase.  We first determine
the hexatic $\to$ isotropic fluid transition temperature $T_i$ and
then check if Frank's constant takes the value $72/\pi$ at $T_i$,
thus testing the KTHNY theory and its prediction that disclination
unbinding occurs at $T_i$. In addition, we analyze the divergence
behavior of the orientational correlation length at $T_i$ and of
Frank's constant at $T_m$.

Different theoretical approaches invoking grain boundary induced
melting \cite{Chui83,Kleinert83} or condensation of geometrical
defects \cite{Glaser93,Lansac06} suggest one first order
transition. However some simulations for Lennard-Jones systems
indicate the hexatic phase to be metastable \cite{Chen95,Somer97}.
The transition in hard-core systems seem to be first-order
\cite{Jaster99} probably due to finite-size effects \cite{Mak06}.
Simulations with long-range dipole-dipole interaction clearly show
second order behavior \cite{Lin06}. Experimental evidence for the
hexatic phase has been demonstrated for colloidal systems
\cite{Murray87,Tang89,Kusner94,Marcus96,Zahn00,Petukhov05}, in
block copolymer films \cite{Segalman03,Angelescu05}, as well as
for magnetic bubble arrays and macroscopic granular or atomic
systems \cite{Seshadi92,Reis06,Zheng06,Dimon85,Li05}. Still the
order of the transitions is seen to be inconsistent. The
observation of a phase equilibrium isotropic/hexatic
\cite{Marcus96,Angelescu05} and hexatic/crystalline
\cite{Marcus96} indicates two first order transitions. In our
system we find two continuous transitions.

\begin{figure*}
\includegraphics[width=0.85 \textwidth]{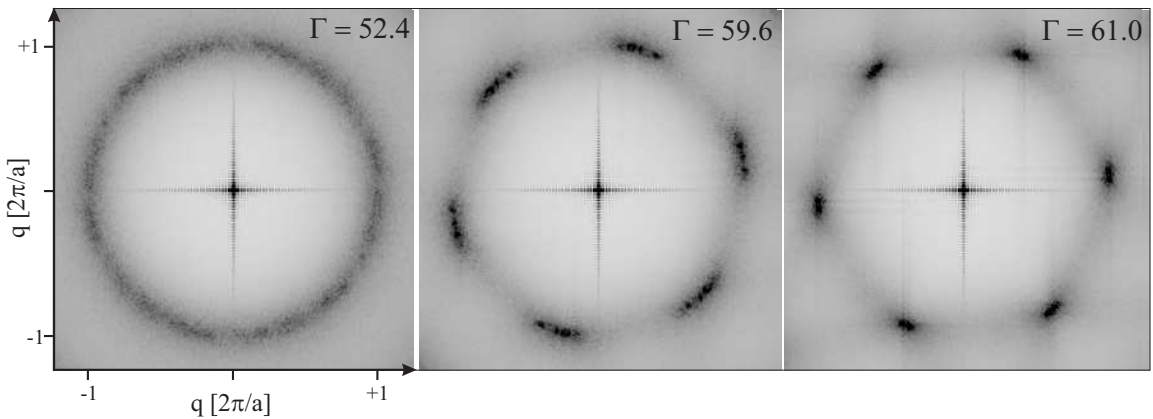}
\caption{\label{fig1}
  Structure factor $ S ( \vec q )$ of our colloidal
  system at three different inverse temperatures $\Gamma$
  corresponding to the isotropic liquid ($\Gamma = 52.4 $), the
  hexatic phase ($\Gamma = 59.6 $) and the crystalline ($\Gamma = 61.0
  $) phase. The central cross is an artifact of the
  Fourier-transformation.}
\end{figure*}

The experimental setup is essentially the same as in
\cite{Keim04}. Spherical and super-paramagnetic colloids (diameter
$d=4.5 \:\mu m$) are confined by gravity to a water/air interface
formed by a water drop suspended by surface tension in a top
sealed cylindrical hole of a glass plate. The field of view has a
size of $835 \times 620 \: \mu m^2$ containing typically up to
$3\cdot10^{3}$ particles (out of $3*10^{5}$ of the whole sample).
A magnetic field $\vec{H}$ is applied perpendicular to the
air/water interface inducing in each particle a magnetic moment
$\vec{M}= \chi \vec{H}$. This leads to a repulsive dipole-dipole
pair-interaction with the dimensionless interaction strength given
by $\Gamma = \beta (\mu_{0}/4 \pi) (\chi H)^{2} (\pi \rho)^{3/2}$.
Here $\chi$ is the susceptibility per colloid while $\rho$ is the
2d particle density and the average particle distance is
$a=1/\sqrt\rho$. The interaction strength can be externally
controlled by means of the magnetic field $H$; it can be
interpreted as an inverse temperature and is the only parameter
controlling the phase behavior of the system. For each $\Gamma$
the coordinates of the colloids are recorded via video-microscopy
(resolution of particle position $dr = 100~nm$) and digital image
processing over a period of $1-2~h$ using a frame rate of
$250~ms$.\\

To set the stage we first visualize in Fig.~(\ref{fig1}) the three
phases and their symmetries by plotting the structure factor
\begin{equation}
\label{sq}
 S(\vec{q}) = \frac{1}{N}\langle\sum_{\alpha,\alpha'}
e^{-i\vec{q}(\vec{r}_\alpha-\vec{r}_{\alpha'})}\rangle\quad ,
\end{equation}
as calculated from the positional data of the colloids for three
different temperatures. Here, $\alpha ,\alpha'$ runs over all $N$
particles in the field of view while $\langle \rangle$ denote the
time average over $700$ configurations.  In the liquid phase,
concentric rings appear having radii that can be connected to
typical inter-particle distances. The hexatic phase, on the other
hand, is characterized by six segments of a ring which arise due
to the quasi long-range orientational order of the six-fold
director \cite{rem1}. In the crystalline phase the Bragg peaks of
a hexagonal crystal show up with a finite width that is due to the
quasi long-range character of the translational order.

\begin{figure}[b]
\includegraphics[width=0.48 \textwidth]{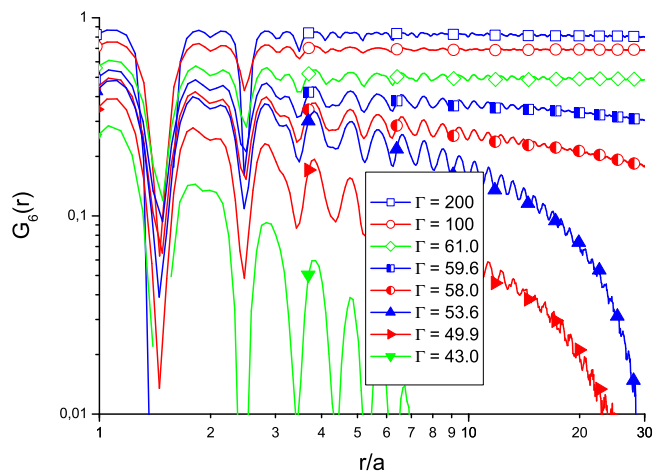}
\caption{\label{fig2} Orientational correlation function $G_6(r)$ as
function of the inverse temperature $\Gamma$ in a log-log plot.  From
top to bottom: Three curves for the crystalline phase showing the
long-range orientational order ($lim_{r\rightarrow\infty} G_6(r) \neq
0$), two curves showing the quasi long-range order of the hexatic
phase ($G_6(r) \sim r^{-\eta_6(\Gamma)}$) and three curves showing the
short-range order typical of the isotropic liquid ($G_6(r) \sim e^{-r
/\xi_6(\Gamma)}$).  }
\end{figure}

To quantify the six-fold orientational symmetry the bond-order
correlation function
\begin{equation}
\label{g6} G_6(r) = \langle \psi(\vec{r})\psi^\ast(\vec{0})
\rangle\quad ,
\end{equation}
is calculated with $\psi(\vec{r}) =
\frac{1}{N_j}\sum_{j}{e^{6i\theta_{ij}(\vec{r})}}$. Here the sum
runs over the $N_j$ next neighbors of the particle $i$ at position
$\vec{r}$ and $\theta_{ij}(\vec{r})$ is the angle between a fixed
reference axis and the bond of the particle $i$ and its neighbor
$j$. $\langle \rangle$ here denotes not only the ensemble average
which is taken over all $N(N-1)/2$ particle-pair distances for
each configuration (resolution $dr = 100~nm$) but also the time
average over $70$ statistically independent configurations. KTHNY
theory predicts that
\begin{eqnarray}
\label{3g6} lim_{r\to\infty} &G_6(r)& \neq 0 \quad\quad\quad
\mbox{crystal: long range order}\nonumber\\
&G_6(r)& \sim r^{-\eta_6} ~~\quad\mbox{hexatic: quasi long range}\nonumber\\
&G_6(r)& \sim e^{-r/\xi_6} \quad\mbox{isotropic: short
range}\nonumber\quad ,
\end{eqnarray}
$\eta_6<1/4$ and takes the value $1/4$ right at $T=T_i$. All three
regimes can be easily distinguished in Fig.~(\ref{fig2}) showing
$G_6(r)$ for a few representative temperatures. Note, that
$G_6(0)$ is not normalized to $1$.

\begin{figure}[t]
\includegraphics[width=0.48 \textwidth]{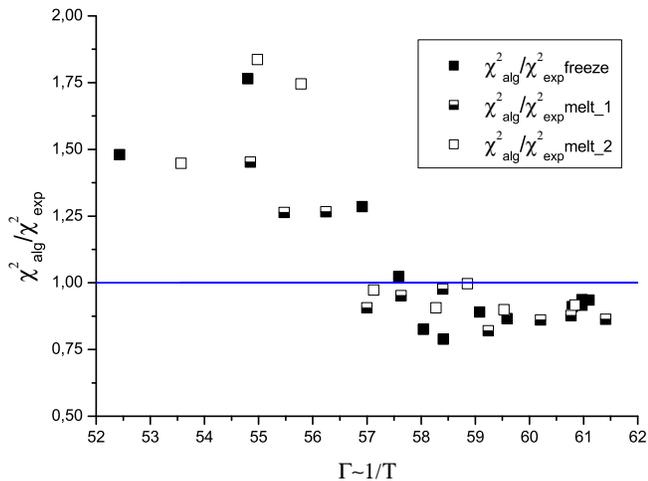}
\caption{\label{fig3}
  Quantitative test for the long distance behavior of $G_6(r)$.
  For $\chi^2_{alg}/\chi^2_{exp}<1$ the algebraic decay fits better.}
\end{figure}

We next fit $G_6(r)$ to $r^{-\eta_6}$ and $e^{-r/\xi_6}$ to
extract $\eta_6$ and $\xi_6$. The fits are performed for radii
$r/a \in \{0..20\}$ \cite{rem2}. To check for the characteristics
of the orientational correlation function, the ratio of the
reduced chi-square $\chi^2$ goodness-of-fit statistic of the
algebraic ($\chi^2_{alg}$) and exponential ($\chi^2_{exp}$) fit is
shown in Fig.~(\ref{fig3}) as a function of $\Gamma$ for three
different measurements. For melting, a crystal free of
dislocations was grown at high $\Gamma$ and then $\Gamma$ was
reduced in small steps. For each temperature step the system was
equilibrated $1/2~h$ before data acquisition started. This was
done at different densities: $melt\_1$ with average particle
distance of $a=11.8~\mu m$ and $melt\_2$ with $a=14.8~\mu m$
containing $3200$ respectively $2000$ particles in the field of
view. The measurement denoted $freeze$ in Fig.~(\ref{fig3})
($a=11.8~\mu m$) started in the isotropic liquid phase and
$\Gamma$ was increased with an equilibration time of $1~h$ between
the steps. For $\chi^2_{alg}/\chi^2_{exp} > 1$ an exponential
decay fits better than the algebraic and vice versa for
$\chi^2_{alg}/\chi^2_{exp} < 1$. We observe in Fig.~(\ref{fig3})
that the change in the characteristic appears at $\Gamma_i=57.5
\pm 0.5$. This value is the temperature of the hexatic
$\leftrightarrow$ isotropic liquid transition.

\begin{figure}[!b]
\includegraphics[width=0.48 \textwidth]{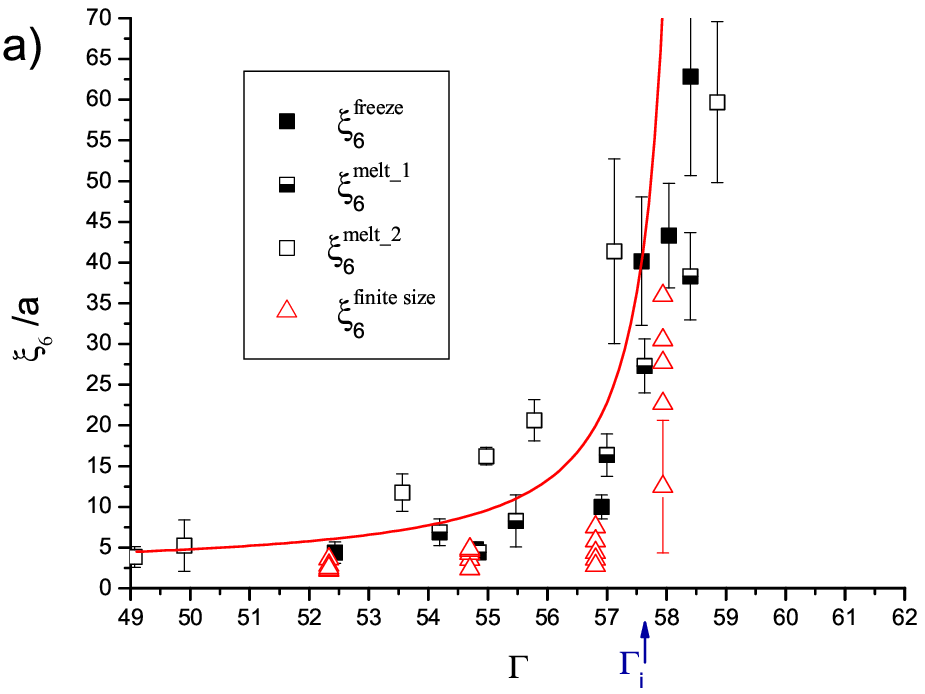}
\includegraphics[width=0.48 \textwidth]{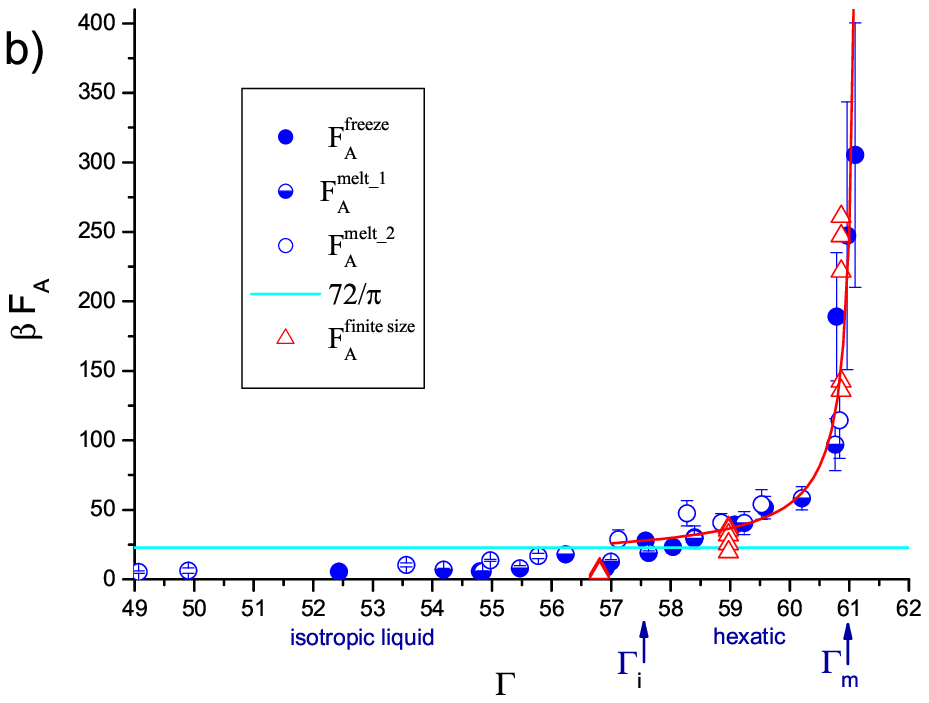}
\caption{\label{fig4} Correlation length $\xi_6$ (a) and Frank
constant $F_A$ (b) as a function of the inverse temperature.
$\xi_6$ diverges at $\Gamma_i$ and $F_A$ at $\Gamma_m$.  In
between the system shows hexatic symmetry. The solid lines are
fits to eq.~(\ref{divxi6}) and (\ref{divFa}) resulting in critical
exponents $\nu = 0.5\pm 0.03$ and $\bar{\nu} = 0.35\pm 0.02$
respectively. Triangles are shifted by $0.1~\Gamma$ for clarity.}
\end{figure}

In the vicinity of the phase transition, approaching $\Gamma_i$
from the isotropic liquid the orientational correlation length
$\xi_6$ should diverge as \cite{Nelson79},
\begin{equation}
\label{divxi6}
  \xi_6(\Gamma) \sim
  \exp~(\frac{b}{|1/\Gamma-1/\Gamma_i|^{\nu}})\quad ,
\end{equation}
with $b$ a constant and $\nu=1/2$. This behavior is observed in
Fig.~(\ref{fig4}a). $\xi_6$ indeed increases dramatically near
$\Gamma_i=57.5 \pm 0.5$ irrespective of whether the system is
heated or cooled. Before discussing this feature we first address
the finite size effect. To this end, we have computed $G_6(r)$ and
$\xi_6$ for subsystems of different size ranging from $720\times
515~\mu m^2,~ 615\times 405~\mu m^2,~ 505\times 300~\mu m^2,~
400\times 190~\mu m^2$ to $ 390\times 80~\mu m^2$. The resulting
data-points are plotted as triangles in Fig.~(\ref{fig4}) and
belong to the black filled squares which they converge to. No
finite size effect is found for $\Gamma<56$, but a considerable
one at $\Gamma=56.9$ close to $\Gamma_i$ where we obviously need
the full field of view to capture the characteristic of the
divergence.  At $\Gamma=58.0$ there is a huge finite size effect
indicating that $\xi_6$ is much larger than the field of view.
However, inside the hexatic phase, $\xi_6$ is no longer well
defined as the decay is algebraic. We fit our data to
eq.~(\ref{divxi6}) in the range $49 < \Gamma < 57.5$ and find the
critical exponent $\nu = 0.5 \pm 0.03$ and $\Gamma_i = 58.9 \pm
1.1$, a value which due to the finite-size effect is slightly
larger than $\Gamma_i$ obtained from Fig.~(\ref{fig3}).

The exponent $\eta_6$ is related to Frank's constant $F_A$
\cite{Nelson79}
\begin{equation}
\label{FA}
 \eta_6(\Gamma) = \frac{18 k_B T}{\pi F_A(\Gamma)}\quad .
\end{equation}
So the critical exponent $\eta_6(\Gamma_i) = 1/4$ corresponds to
$\beta F_A(\Gamma_i) = 72/ \pi$ at the hexatic $\leftrightarrow$
liquid transition. This quantity is plotted in Fig.~(\ref{fig4}b).
Indeed, $F_A$ crosses the value $72/\pi$ at $\Gamma_i=57.5 \pm 0.5$
exactly at that temperature which in Fig.~(\ref{fig3}) has been
independently determined to be the transition temperature $T_i$.  For
$\Gamma < \Gamma_i$, $F_A$ should jump to zero which is not completely
reproduced. We note that since $\eta_6$ is not well defined in the
isotropic fluid, it becomes problematic to extract $F_A$ from
eqn. (\ref{FA}) below $\Gamma_i$.  At $\Gamma_m$, at the hexatic
$\rightarrow$ crystalline transition, $F_A$ must diverge which indeed
it does.  This divergence can be identified with the divergence of the
square of the translational correlation length $\xi_+$ \cite{Nelson79}
\begin{equation}
\label{divFa} F_A(\Gamma)/k_B T \sim \xi_+^2 \sim
\exp~(\frac{2c}{|1/\Gamma-1/\Gamma_m|^{\bar{\nu}}})\quad ,
\end{equation}
where $c$ is again a constant and $\bar{\nu}=0.36963$. Fitting the
values of $F_A$ to the expression in eqn.~(\ref{divFa}) in the
range $57.5 < \Gamma < 61$ we obtain $\bar{\nu} = 0.35 \pm 0.02$
and $\Gamma_m = 61.3 \pm 0.4$ as an upper threshold. Again
triangles represent evaluation of our data in sub-windows of
variable size (same sizes as above). The finite size effect for
$\Gamma = 57.0$ is negligible. Close to $\Gamma_m$ it increases
but the values saturates for $\Gamma = 59.1$ and $\Gamma = 60.8$
and remain within the error-bars for the biggest sub-windows.\\

In conclusion, we have checked quantitatively the change of
quasi-long-range to short-range orientational order and extracted
the correlation length $\xi_6$ in the isotropic fluid and Frank's
constant $F_A$ in the hexatic phase from trajectories of a 2d
colloidal system. We find a hexatic $\leftrightarrow$ isotropic
liquid transition at $\Gamma_i=57.5 \pm 0.5$. Three observations
support this result: (i) the change of the distance dependence of
$G_6(r)$ (Fig.~(\ref{fig3})), (ii) the condition $F_A(\Gamma_i) =
72/\pi$ for Frank's constant and (iii) the divergence of $\xi_6$.
For the transition hexatic $\leftrightarrow$ crystal $F_A$
diverges at $\Gamma_m$. Both divergencies (extracted from just one
correlation function) lead to critical exponents that are in good
agreement with the KTHNY-theory. The measurements for melting and
freezing support each other; so we may conclude that there is no
hysteresis effect of the phase-transitions. At the two
transitions, the order parameters are observed to change
continuously (within the resolution of $\Gamma \propto 1/T$); no
indication of a phase-separation (as for example strong
fluctuations of the order parameters) has been found \cite{rem3}
as has been reported by [17, 21]. So we believe that in our system
- having a well-defined, purely repulsive pair-potential and a
confinement to 2D that is free of any surface roughness - the
transitions are second order.

In \cite{Gruenberg04,Zanghellini05} we verified that the Young's
modulus becomes $16\pi$ at $T_m$. We have now checked that $F_A$
takes the value $72/\pi$ at $T_i$. These two findings together
confirm the two-stage KTHNY melting scenario with its underlying
microscopic picture of breaking the translational symmetry by
dislocation-pair- and orientational symmetry by
disclination-pair-unbinding.\\

P. Keim gratefully acknowledges the financial support of the
Deutsche Forschungsgemeinschaft.


\begin{thebibliography}{10}
\bibitem{rem1} The segments will merge to rings, if the systems size tends towards infinity.
\bibitem{rem2} The upper value is motivated by the maximum value of
the histogram of distances and $r/a=0$ is excluded in the
algebraic case to avoid the singularity. The histrogram over
distances is used as statistical weight of the fit-function taking
into account the different frequency of occurrence of the
data-points in the minima and maxima of $G_6(r)$.
\bibitem{rem3} We indeed do see a local clustering of dislocations as already noted
by \cite{Murray87,Tang89} in the hexatic phase close to $\Gamma_i$
(which is not too surprising for dislocations with finite density
having an attractive interaction) but this averages out in $G_6$
if the field of view is big enough.
\end{thebibliography}
\end{document}